# Data-driven soliton solutions and model parameters of nonlinear wave models via the conservation-law constrained neural network method

Yin Fang, Gang-Zhou Wu, Yue-Yue Wang [*] and Chao-Qing Dai [*]

*College of Optical, Mechanical and Electrical Engineering, Zhejiang A&F University, Lin'an, Zhejiang 311300, P. R. China*

**Abstract.** In the process of the deep learning, we integrate more integrable information of nonlinear wave models, such as the conservation law obtained from the integrable theory, into the neural network structure, and propose a conservation-law constrained neural network method with the flexible learning rate to predict solutions and parameters of nonlinear wave models. As some examples, we study real and complex typical nonlinear wave models, including nonlinear Schrödinger equation, Korteweg-de Vries and modified Korteweg-de Vries equations. Compared with the traditional physics-informed neural network method, this new method can more accurately predict solutions and parameters of some specific nonlinear wave models even when less information is needed, for example, in the absence of the boundary conditions. This provides a reference to further study solutions of nonlinear wave models by combining the deep learning and the integrable theory.

## 1. Introduction

Soliton theory has been deeply studied and widely used in mechanics, physics, biology, hydrodynamics and other engineering fields [1-3], and the essence of soliton is a special wave-like solutions of nonlinear wave models, namely nonlinear partial differential equations (NPDEs) [4]. When the nonlinear term and dispersion term in these NPDEs are balanced, a stable solitary wave solution can be formed [5]. The integrable NPDEs with soliton solutions have become the key research objects in the fields of applied mechanics and engineering, including some typical nonlinear wave models, such as nonlinear Schrödinger equation (NLSE), Korteweg-de Vries (KdV) equation, modified Korteweg-de Vries (mKdV) equation and so on. Moreover, these integrable NPDEs have many invariant properties. Noether [6] theorem points out that there is a specific invariance, and there is a corresponding conservation law. Space invariance corresponds to the law of conservation of momentum, and time translation invariance corresponds to the law of conservation of energy [7,8]. Most of NPDEs with these conservation laws are completely integrable nonlinear wave models.

---
* Corresponding author email：wangyy424@163.com (Y.Y. Wang); dcq424@126.com (C.Q. Dai)

The study of soliton solutions of NPDEs mainly focuses on analytical methods and numerical simulation. However, some equations are difficult to solve by analytical methods. The traditional numerical methods are faced with the problems of high calculation cost and long calculation time. Therefore, it is urgent to introduce a new method to study NPDEs. In recent years, with the explosive growth of available data and computing resources, the development of artificial neural network has been promoted. Deep learning in the form of deep neural network has been involved in various fields, from image recognition and language translation [9,10] in classification problems to the solution of nonlinear discontinuous functions [11] in regression problems. However, their applications in the field of scientific computing are rarely studied. Researchers found that neural network has a strong ability to deal with strong nonlinear and high-dimensional problems [12]. It has been shown that deep neural network, as a general function approximator [13], can overcome the dimension disaster in some problems [14-16], which also makes neural network have a bright application prospect in the field of scientific computing and become a popular alternative modeling method.

The problem of traditional machine learning can only learn from the marked data, but can not know the mechanism behind the system. On the contrary, for the modeling of physical systems, the governing equations are usually known, but it is difficult to solve them effectively. As early as the late 1990s, it was proposed to use the states at some points in the parameter space and combined with the known governing equations to restrict (or even drive) learning, which can make up for the lack of data [17,18]. However, limited by the neural network technology and computing power at that time, this pioneering work did not have much impact.

Recently, Raissi et al. proposed a physics-informed neural network (PINN) to solve NPDEs [19]. Due to the addition of NPDEs as physical information, it also provides a good physical explanation for these predicted solutions. Yan et al. used the PINN method to solve the forward and inverse problems of NLSE with the PT-symmetric harmonic potential [20] and also discussed the data-driven rogue wave solutions of defocusing NLSE [21]. Chen et al. studied the soliton solutions of KdV equation, mKdV equation and KdV-Burgers equation by using the PINN method [22]. Jagtap et al. proposed PINN method of regional conservation on discrete domain to solve Burgers and KdV equations [23].

As we all know, by minimizing the loss of the initial / boundary sampling points and the loss of NPDEs at the configuration points, the PINN method can feed back the optimal parameters of the neural network and the physical parameters of NPDEs. But data with known initial and boundary conditions are required for the PINN method, which is a very stringent requirement. Recent studies relax the requirment of initial and boundary conditions for the PINN method [24,25]. Raissi et al. proposed hidden fluid machine learning, added the Navier-Stokes (NS) equation derived from the conservation laws of mass, momentum and energy to the neural network, and predicted the pressure field and velocity field of NS equation [24]. Sun et al. proposed an alternative modeling method of incompressible fluid with mass conservation and momentum conservation without initial and boundary data to study the forward problem of NS equation [25].

It is obviously not enough to restrict the PINN only by using NPDEs in Refs.[19-23]. More recently, Chen et al. designed a two-stage PINN method to optimize this single restriction in the traditional PINN method. In the second stage, they used the local conserved quantity to constrain the output of PINN, and studied the local wave solutions of classical Boussinesq-burgers equations [26]. However, the conservation laws of NPDEs given by integrable system theory are added to the neural network structure has not been reported to predict soliton solutions of NPDEs.

The conservation laws of NPDEs have corresponding conserved quantities. In the theory of integrable systems, the existence of conserved quantities is closely related to whether the evolution equations of physical systems can be solved by the inverse scattering method. The conservation laws of NPDEs are often used in the analysis of analytical theory, and rarely used in the calculation of numerical algorithms. We propose a conservation-law constrained neural network(CLCNN) method with less data to predict soliton solutions of NPDEs. The novelties of this paper include the following three aspects. (I) Compared with the traditional PINN method, this new CLCNN method can more accurately predict solutions and parameters of some specific NPDEs even when less information and data is needed, for example, in the absence of the boundary conditions; (II) The learning rate of previous neural network algorithms is a fixed parameter, but in this paper, we introduce a learning rate that decreases with the increase of training times, which has a positive effect on the prediction of results; (III) The CLCNN method have a wide range of applications, that is, it has good prediction results for soliton structures of different physical models and different soliton structures of the same physical model.

## 2. Conservation-law constrained neural network method

In most physical systems, the interaction between nonlinearity and dispersion is considered, and the dissipation can be ignored. Therefore, in this paper, we will study time-varying NPDEs, which often play an important role in many scientific applications and physical phenomenons. The specific form of (1+1)-dimensional NPDEs is as follows

$$Q_t = N(Q, Q_x, Q_{xx}, Q_{xxx}). \tag{1}$$

Where subscripts $x$ and $t$ represent the partial derivatives relative to space and time, and $N$ represents the nonlinear function combination of solution $Q$ and its arbitrary partial derivatives relative to space variable $x$. Specifically, we utilize the depth neural network to approximate the solution of Eq. (1), and then calculate the derivative of $Q$ relative to space $x$ and time $t$ with the help of automatic differentiation network. Therefore, the NPDE is defined as

$$f := Q_t - N(Q, Q_x, Q_{xx}, Q_{xxx}), \tag{2}$$

which is a complex equation. After separating the real and imaginary parts, we can get

$$f_r := r_t - N(r, r_x, r_{xx}, m, m_x, m_{xx}, \ldots),$$
$$f_m := m_t + N(r, r_x, r_{xx}, m, m_x, m_{xx}, \ldots),$$
(3)

where $r, m$ represent the real and imaginary parts of solution $Q$ respectively. The neural network learns shared parameters (such as weights and deviations) by minimizing the mean square error (MSE) sum caused by the initial conditions, NPDEs, and conservation laws related to the feedforward neural network. The specific form of the loss function is as follows

$$loss = MSE_0 + MSE_c + MSE_h,$$
(4)

in which

$$MSE_0 = \frac{1}{N_0} \sum_{i=1}^{N_0} (|r(x^i, t^i) - r^i|^2 + |m(x^i, t^i) - m^i|^2),$$

$$MSE_c = \frac{1}{N_f} \sum_{j=1}^{N_f} [a(|f_{ECr}(x^j, t^j)|^2 + |f_{ECm}(x^j, t^j)|^2) + b(|f_{MCr}(x^j, t^j)|^2 + |f_{MCm}(x^j, t^j)|^2)],$$
(5)

$$MSE_h = \frac{1}{N_f} \sum_{j=1}^{N_f} (|f_r(x^j, t^j)|^2 + |f_m(x^j, t^j)|^2).$$

Here $\{r^i, m^i\}_{i=1}^{N_0}$ represent the sampling points on the mean square error caused by the initial conditions of solution $Q$, $\{r^j, m^j\}_{j=1}^{N_f}$ denote the configuration points on the mean square error $MSE_h$ caused by the NPDE and the mean square error $MSE_c$ caused by the conservation laws derived from integrability of NPDEs in soliton theory. $f_{ECr}, f_{ECm}$ and $f_{MCr}, f_{MCm}$ represent the real and imaginary parts of the mean square error caused by the energy and momentum conservation laws of the NPDE, respectively. In this article, we choose $N_0 = 50, N_f = 10000, a = 0,1, b = 0,1,$ the activation function is the adaptive sin function and the optimizer is Adam. At the same time, we set the learning rate of the optimizer to a flexible learning rate, and its size becomes 90% of the original size every 300 training sessions. The schematic diagram of the CLCNN method is given in Fig. 1.

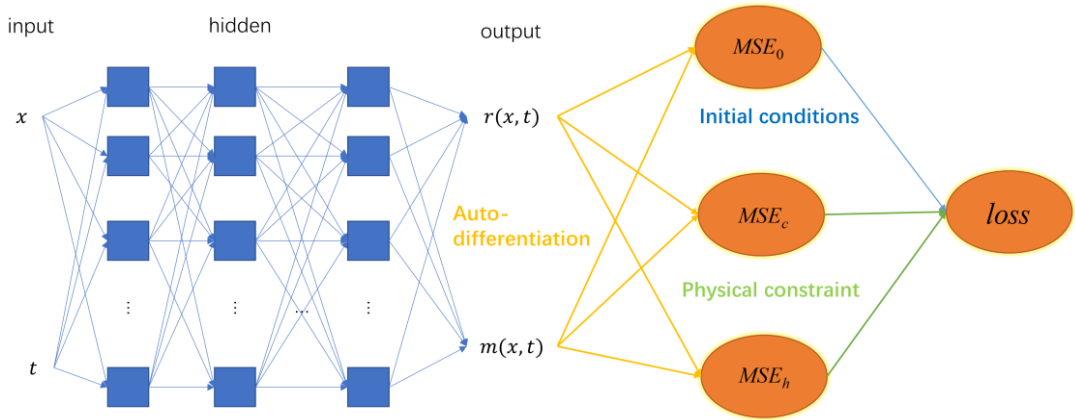

**Fig.1.** The schematic diagram of the CLCNN method. Here $MSE_o, MSE_c$ and $MSE_h$ denote mean square errors caused by initial conditions, conservation laws and NPDE, respectively.

If Eq. (2) is a real solution equation, the output of the CLCNN become an output, and the mean square error of the imaginary part in the loss function is 0, the complex solution algorithm degenerates to a real solution algorithm.

The integrability in soliton theory gives an infinite number of conservation laws for NPDEs. At present, some conservation laws related to the integrability in soliton theory are found to have physical explanations, such as the conservation of energy, the conservation of momentum, or their combination, which are introduced into the loss function to replace the residual of the boundary condition in the CLCNN method. Our research indicates that the CLCNN method has good performance on the prediction of solutions and parameters for NLSE, KdV and mKdV equations, which also hints that this method has universality and generalization.

## 3. Prediction of soliton solutions of NLSE

In fluid mechanics, NLSE can describe the formation of vortices and deep water waves [27]. In optics, the propagation of picosecond optical soliton in a single-mode fiber is controlled by the standard NLSE, which is a fully integrable NPDE. This equation has rich soliton solutions, and its specific form is as follows

$$iQ_t + Q_{xx} + 2|Q|^2 Q = 0. \tag{6}$$

In nonlinear optics [28], $Q$ represents the pulse slowly varying amplitude envelope, and $x, t$ represent the normalized distance and time coordinates. In Bose Einstein condensates [29], $Q$ represents the order parameter and $t, x$ represent the time and space coordinates.

According to the integrable theory, starting from the standard NLSE, we use the Lax pair to construct the corresponding Darboux transformation, and then can deduce its energy conservation and momentum conservation laws [30]. Separating their real and imaginary parts yields.

$$\begin{aligned}
f_{ECr} &:= rr_t + mm_t - r_{xx}m + m_{xx}r, \\
f_{ECm} &:= 0, \\
f_{MCr} &:= r_t r_x + m_t m_x + rr_{xt} + mm_{xt} - r_{xx}m_x + m_{xx}r_x + rm_{xxx} - mr_{xxx}, \\
f_{MCm} &:= m_t r_x - r_t m_x - rm_{xt} + mr_{xt} - r_{xx}r_x - m_{xx}m_x + rr_{xxx} + mm_{xxx} + 4r^3 r_x + 4m^3 m_x + 4rr_x m^2 + 4mm_x r^2.
\end{aligned} \tag{7}$$

Next, we will consider the following combinations of conservation laws constraint to train the neural network.

$$\begin{aligned}
L1 &= Eq + MC + EC + initial, \\
L2 &= MC + EC + initial, \\
L3 &= Eq + MC + initial, \\
L4 &= Eq + EC + initial, \\
L5 &= Eq + initial + boundary.
\end{aligned} \tag{8}$$

where $MC, EC, Eq$ represent momentum conservation law, energy conservation law and NLSE respectively. $L5$ is the loss function combination used in the traditional PINN method. In all examples of NLSE, in terms of the relative error of the real and imaginary parts of network prediction, we find that the best combination is the form $L4$, that is, the combination of the original equation and the energy conservation law. The loss for this specific combination is as follows

$$loss = \frac{1}{N_0}\sum_{i=1}^{N_0}(|r(x^i,t^i)-r^i|^2 + |m(x^i,t^i)-m^i|^2) + \frac{1}{N_f}\sum_{j=1}^{N_f}(|f_{ECr}(x^j,t^j)|^2 + |f_{ECm}(x^j,t^j)|^2 + |f_r(x^j,t^j)|^2 + |f_m(x^j,t^j)|^2). \quad (9)$$

### 3.1. One-soliton solution

The exact one-soliton solution [31] of Eq. (6) reads

$$Q(x,t) = 0.6\,\text{sech}(0.6x)e^{i}, x \in [-15,15], t \in [0,3]. \quad (10)$$

From the known range of local spatiotemporal region $x,t$, we can obtain the initial and boundary conditions of the equation. The data set can be obtained by pseudo spectral method, and the exact one-soliton solution can be dissociated into data points. The corresponding number of sampling points is given in the section 2.

Table.1. Comparison of prediction results with different types of learning rates for one-soliton

| Learning rate type | Loss | Relative error of $Q(x,t)$ |
| --- | --- | --- |
| Fixed learning rate | $3.501\times10^{-6}$ | $8.776445\times10^{-3}$ |
| Flexible learning rate | $2.000\times10^{-6}$ | $3.670414\times10^{-3}$ |

The learning rate is a relatively important parameter of neural network. In the past, when the neural network method were used to solve NPDE, the learning rate was always a fixed value, which would make the loss function fluctuate greatly during the training process, and the network optimization is extremely easy falling into a local minimum, and thus is not conducive to prediction. To solve the above problems, we introduced a variable learning rate, that is, the initial learning rate is 0.001, and the learning rate becomes 90% of the original size after every 200 steps of training. Next, we use the one-soliton solution of the NLSE as a model to discuss the influence of this flexible learning rate on the convergence speed and convergence size for the loss function combination $L5$ used in the traditional PINN method. From Fig. 2, we find that the flexible learning rate makes the loss curve smoother, the convergence speed is faster, and the convergence size is small. Table 1 shows that the flexible learning rate makes relative error for $Q(x,t)$ of the network prediction smaller. Here the relative error is $[|\hat{Q}(x,t)-Q(x,t)|]/Q(x,t)$ with exact solution $Q(x,t)$ and predicted solution $Q(x,t)$. Therefore, we use this flexible form of learning rate in this article.

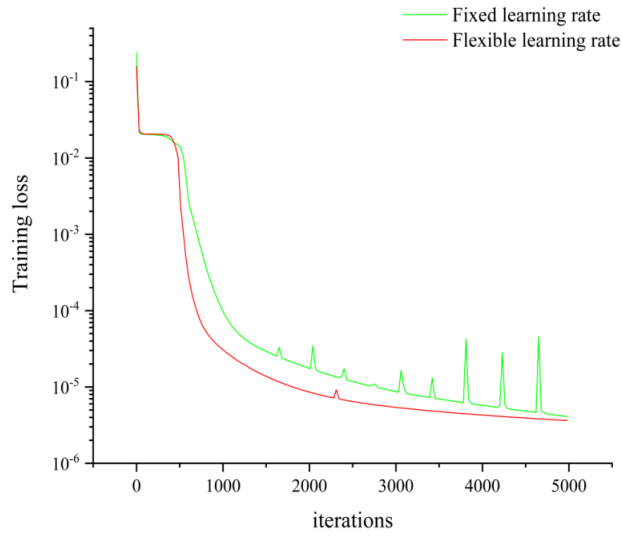

**Fig.2.** The impact of different types of learning rates on the loss function

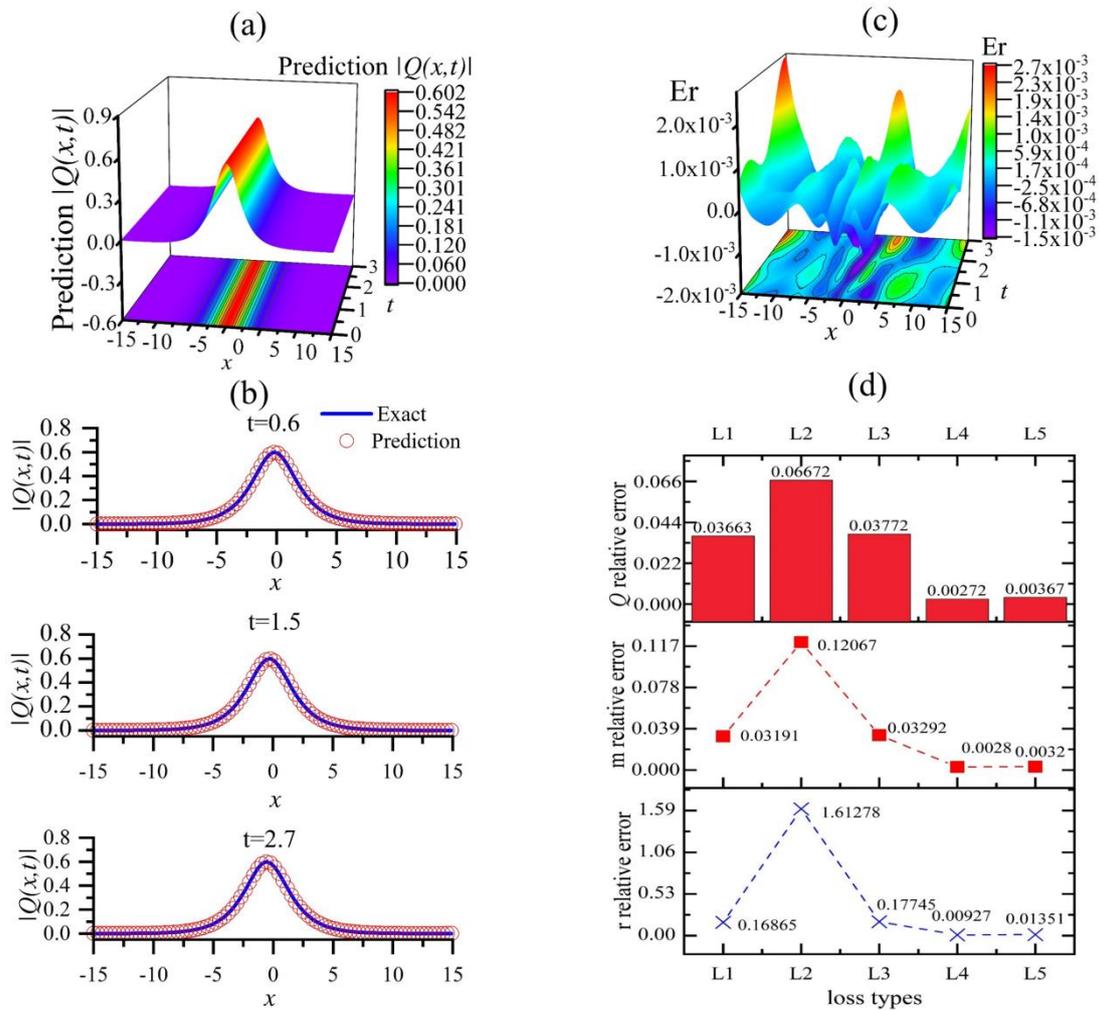

**Fig.3.** One-soliton solution $Q(x,t)$ of (10). (a) Predicted result of one-soliton, (b) comparison between exact and predicted solutions at different propagation distances, (c) diagram of error between exact and predicted solutions, and (d) relative errors of the predicted solution for different

loss combinations.

We use the CLCNN method to predict the dynamic behavior of one-soliton for NLSE. In the training process, the Latin sampling [32] method is used to only obtain the initial configuration points, and the data of the boundary condition are not used. Fig. 3(a) shows the dynamic behavior of one-soliton solution reconstructed by the CLCNN method. Fig. 3(b) shows the detailed comparison between the predicted solution and exact solution at the distances $t=0.6, t=1.5, t=2.7$. It is found that the predicted result fits well with the evolution described by exact solution. Fig. 3(c) shows the gap $Er = Q(x,t) - \hat{Q}(x,t)$ between the exact solution $Q(x,t)$ and predicted solution $\hat{Q}(x,t)$, we find that the gap between the exact and predicted solutions is very small, reaching level $10^{-3}$. Fig. 3(d) presents predicted results of NLSE with four different combinations of conservation laws and compares them with result by the loss function combination $L5$ used in the traditional PINN method.

In Fig. 3(d), the upper, middle and lower figures exhibit the relative errors of one-soliton solution and its imaginary and real parts respectively. The relative errors of imaginary and real parts is $[\|\hat{m}(t,x) - m(t,x)\|]/m(t,x), [\|\hat{r}(t,x) - r(t,x)\|]/r(t,x)$ respectively. From Fig. 3(d), the relative error for the loss combination $L4$ of NLSE, the energy conservation law and initial conditions is the smallest, while that for the loss combination $L2$ of the energy conservation law, momentum conservation law and initial conditions is the largest. The reason for this is that the lack of a dynamic model will lose important physical information, which makes it difficult for the neural network to predict an excellent solution.

### 3.2. Two-soliton solution

Exact solution two-soliton [31] of Eq. (6) is

$$Q(x,t) = \frac{-2i(-0.246ie^{0.64it+1.4x} + 0.462ie^{1.96it-0.8x} - 0.264ie^{0.64it-1.4x} + 0.462ie^{1.96it+0.8x})}{-1.12e^{-1.32it} - 1.12e^{1.32it} + 1.21e^{-0.6x} + 1.21e^{0.6x} + 0.09e^{-2.2x} + 0.09e^{2.2x}}, \quad \text{(Interaction)},$$

$$Q(x,t) = \frac{-2i(-0.05ie^{0.7744it-0.8x} - 0.06ie^{0.7744it+0.8x} - 0.05ie^{0.64it+0.88x} - 0.05ie^{0.64it-0.88x})}{-1.32e^{-0.1344it} - 1.32e^{0.1344it} - 1.41e^{-0.08x} - 1.23e^{0.08x} + 0.0004e^{-1.68x} + 0.0037e^{1.68x}}, \quad \text{(molecule)}.$$

(11)

From the scope of the local space-time region of the interaction between two solitons as $x \in [-8,8], t \in [-2,2]$, and the scope of the space-time region of the soliton molecule as $x \in [-15,15], t \in [0,3]$, we get the initial and boundary conditions of NLSE. Other operation is similar to one-soliton solution in the section 3.1.

Table.2. Relative errors of two-soliton solution with different types of loss

| Soliton type | relative error | Loss type | | | | |
|---|---|---|---|---|---|---|
| | | L1 | L2 | L3 | L4 | L5 |
| Soliton interaction | r | 0.115076 | 1.356908 | 0.125183 | 0.071768 | 0.077711 |
| | m | 0.255874 | 1.663534 | 0.289671 | 0.159936 | 0.178356 |
| | Q | 0.068093 | 0.145672 | 0.064015 | 0.037836 | 0.047003 |
| Soliton | r | 0.079548 | 1.184964 | 0.088480 | 0.068112 | 0.137107 |
| molecule | $m$ | 0.034496 | 0.938924 | 0.038073 | 0.029569 | 0.062584 |
| | $Q$ | 0.021590 | 0.041609 | 0.024117 | 0.018582 | 0.039642 |

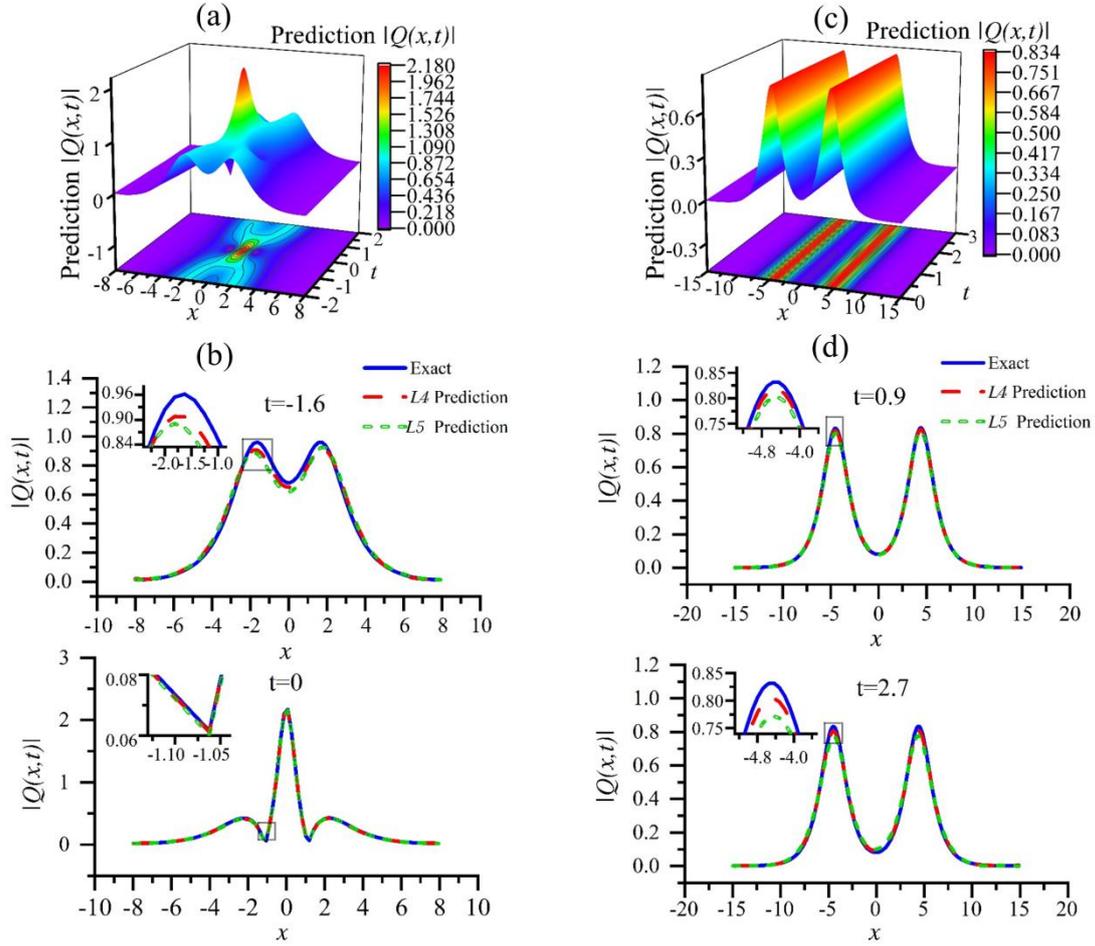

**Fig.4.** Two-soliton solution $Q(x,t)$ of (11). Predicted results of (a) two interacting solitons and (c) two-soliton molecule, comparison between exact and predicted solutions of (b) two interacting solitons and (d) two-soliton molecule at different propagation distances. The blue solid line is exact solution, the red dotted line and green dash line denote the predicted solutions with the combinations $L4$ and $L5$ respectively.

In the training process, the Latin sampling method is used to only obtain the initial configuration points, and the data of the boundary condition are not used. Table 2 presents the predicting results of two-soliton solution by the CLCNN method with four different combinations of conservation laws. The same conclusion is reached for two interacting solitons and two-soliton molecule, that is, the relative errors of solution and their real and imaginary parts for the loss combination $L4$ of original equation, the energy conservation law and initial conditions are all smallest, while those for the loss combination $L2$ of the energy conservation law, momentum conservation law and initial conditions are all largest. This result is consistent with that of one-soliton solution in the section 3.1.

Figs. 4(a) and 4(c) show dynamic behaviors of two interacting solitons and two-soliton molecule reconstructed by the CLCNN method without boundary information, respectively. Figs. 4(b) and 4(d) show the detailed comparison between the predicted solution and the exact solution at distances $t=-1.6, t=0$ and $t=1.5, t=2.7$, respectively. Compared with results for the loss function combination $L5$ used in the traditional PINN method, we find that predicted results for the loss function combination $L4$ is closer to exact solutions at all distances in Figs. 4(b) and 4(d). From the detailed comparison of these illustrations in Fig. 4(b) and 4(d), as the distance increases, the predicted results for both $L4$ and $L5$ will become worse.

### 3.3. Rogue wave solution

Exact rogue wave solution [31] of Eq. (6) reads

$$Q(x,t)=0.6(1-\frac{4+5.76it}{1+1.44(x-0.0116t)^2+2.0736t^2})e^{-0.71996636i}, x\in[-4,4], t\in[-1.5,1.5]. \quad (12)$$

All operation is similar to one-soliton solution in the section 3.1. In the training process, the Latin sampling method is used to only obtain the initial configuration points, and the data of the boundary condition are not used.

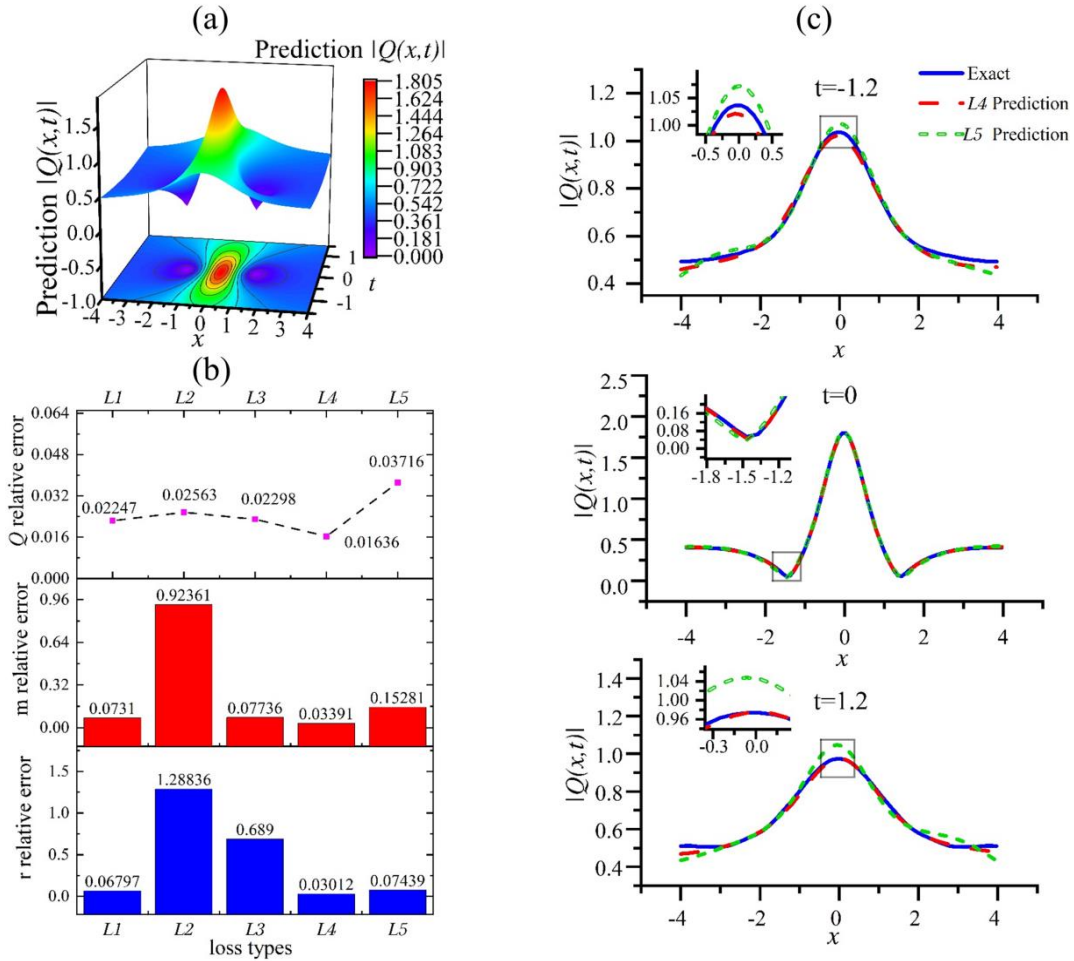

**Fig.5.** Rogue wave solution $Q(x,t)$ of (12). (a) The prediction result, (b) the relative errors of the predicted solutions with different combinations of conservation laws, and (c) comparison between exact solution and predicted solution at different propagation distances. The blue solid line is the exact solution, and the red dotted line and green dash line denote the predicted solution with the combination $L4$ and $L5$ respectively.

In Fig. 5(a), it is shown that the complex local rogue wave solution of the NLSE can be accurately predicted using the CLCNN method without boundary information. Fig. 5(b) shows the relative errors of the real part $r$, imaginary part $m$ and solution $Q$. The result is consistent with that of the previous examples, that is, the relative errors of solution and their real and imaginary parts for the loss combination $L4$ of original equation, the energy conservation law and initial conditions are all smallest, while those for the loss combination $L2$ of the energy conservation law, momentum conservation law and initial conditions are all largest. Fig. 5(c) shows detailed comparison between the predicted solution and exact solution at $t=-1.2, t=0, t=1.2$. Compared with results for the loss function combination $L5$ used in the traditional PINN method, predicted results for the loss function combination $L4$ is closer to exact solutions at all distances in Fig. 5(c).

From the prediction results of the above three local structures, we can see that the CLCNN method without boundary information has good applicability to the complex NLSE, and a suitable loss combination of conservation laws is found, that is, the loss prediction with original equation, energy conservation law and initial condition is the best.

## 4. Prediction of soliton solutions of KdV equation

In the section 3, we have studied the applicability of the the CLCNN method to complex equation. However, whether this method is applicable to real equations, and whether there are similar loss combination of conservation law need to be further elucidated. Next, we take KdV and mKdV equations as examples to study the case of real equations.

The KdV equation is widely used in the fields of physics such as hydrodynamics, plasma, ion acoustic waves, non-resonant lattice vibration [33,34], etc. This equation has abundant soliton solutions. In this section, we will consider the neural network with four different combinations of conservation laws to predict one-soliton and two-soliton solutions of the KdV equation, and compare them with the predicted results of the traditional PINN method.

The form of the KdV equation is as follows

$$Q_t + 6QQ_x + Q_{xxx} = 0, \qquad (13)$$

where $Q$ represents the amplitude of the solitary wave, and the subscript $x, t$ denote the normalized distance and time coordinate.

Starting from Eq. (13), We use the Lax pair of integrable theory to construct the corresponding Darboux transformation, and we can deduce its energy conservation and

momentum conservation laws [35,36,37] as follows

$$f_{EC} := 3Q^2(Q_t + 6QQ_x + Q_{xxx}) + Q_x(Q_{xt} + 6Q_x^2 + 6QQ_{xx} + Q_{xxxx}),$$
$$f_{MC} := Q_{xxt} + 2QQ_t + Q_{xxxx} + 18Q_xQ_{xx} + 8QQ_{xxx} + 12Q^2Q_x. \quad (14)$$

Next, soliton solutions of the KdV equation predicted by $L1, L2, L3, L4, L5$ will be considered. $L1, L2, L3, L4, L5$ types are the same as NLSE in the section 3. In all examples of KdV, we find that the loss function combination of the original equation and the momentum conservation law is best, and the relative error of the prediction result is the smallest, namely L3 form. The loss for this specific combination is as follows

$$loss = \frac{1}{N_0}\sum_{i=1}^{N_0}\left|r(x^i,t^i) - r^i\right|^2 + \frac{1}{N_f}\sum_{j=1}^{N_f}(\left|f_r(x^j,t^j)\right|^2 + \left|f_{MCr}(x^j,t^j)\right|^2). \quad (15)$$

**4.1. One-soliton solution**

The exact one-soliton solution [38] of Eq. (13) is

$$Q(x,t) = \frac{1}{2}\operatorname{sech}^2(\frac{1}{2}(x-t)), x \in [-20, 20], t \in [0,5]. \quad (16)$$

Similar to one-soliton case of the NLSE in the section 3.1, the above exact one soliton is discretized into data points. In the training process, the Latin sampling method is used to only obtain the initial configuration points, and the data of the boundary condition are not used. The corresponding sampling points are given in the section 2.

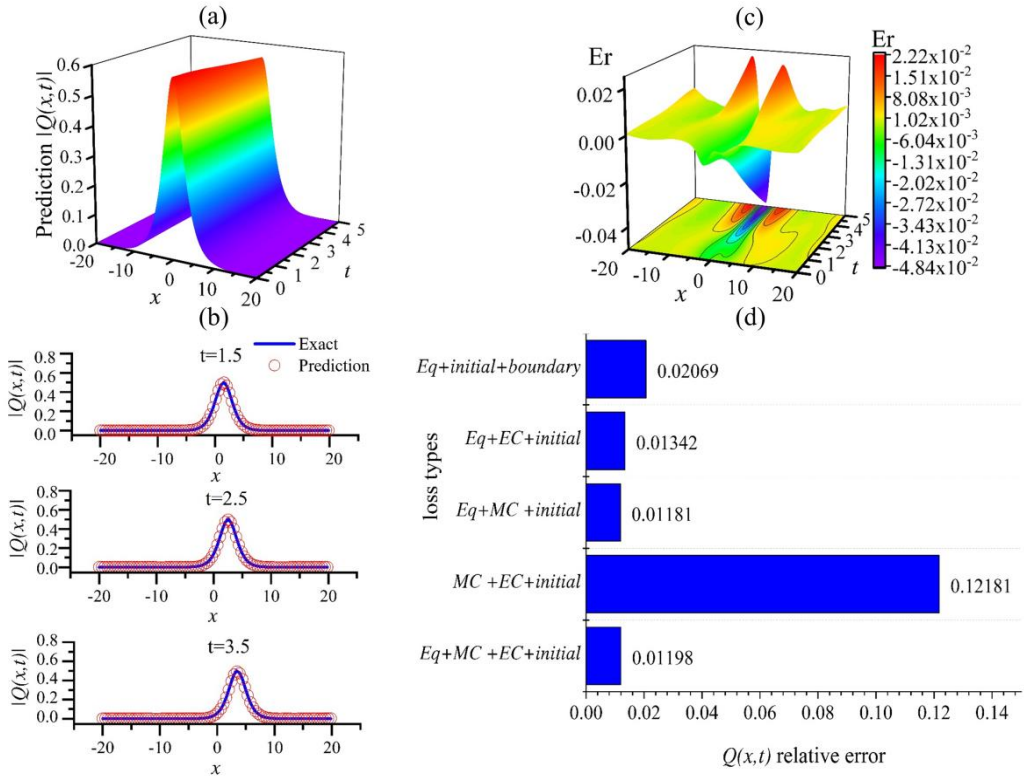

**Fig.6.** One-soliton solution $Q(x,t)$ of (16). (a) Prediction result, (b) comparison between exact and predicted results at different time, (c) diagram of error between exact and predicted solutions, and (d) relative errors of the predicted solution for different loss combinations.

For the one-soliton case of the KdV equation, it is shown that our proposed method is also applicable to real equations. Fig. 6(a) shows the dynamic behavior of one-soliton reconstructed by the CLCNN method without boundary information. Fig. 6(b) shows the detailed comparison between the predicted solution and the exact solution at $t=1.5, t=2.5, t=3.5$. Fig. 6(c) shows shows the gap between the exact solution $Q(x,t)$ and predicted solution $Q(x,t)$, we find that as the transmission distance increases, the error will gradually increase. Fig. 6(d) presents the results of neural network prediction for one-soliton solution with four different combinations of conservation laws, and compares them with result by the loss function combination $L5$ used in the traditional PINN method.

It can be seen from Fig. 6 that the loss function combination including the original equation itself has a good prediction result. When the constraint information contains the original equation and momentum conservation law, the prediction obtains the best result with a relative error of $0.01181331$, which is different from the result of complex NLSE. When the constraint information is the energy and momentum conservation laws, the prediction results are the worst with a relative error is $0.1218068$, which is consistent with the result of complex NLSE.

### 4.2. Two-soliton solution

The exact solution two-soliton [38] of Eq. (13) reads

$$Q(x,t) = \frac{3}{2}\text{sech}^2(\frac{\sqrt{3}}{2}(x-3t)) + \frac{1}{2}\text{sech}^2(\frac{1}{2}(x-t)), x\in[-20,20], t\in[-3,3], \text{(Interaction)}$$
$$Q(x,t) = \frac{1}{2}\text{sech}^2(\frac{1}{2}(x-t+8)) + \frac{1}{2}\text{sech}^2(\frac{1}{2}(x-t-5)), x\in[-20,20], t\in[0,5], \text{(Molecule)}.$$
(17)

From the known range of local spatiotemporal region $x,t$, we can obtain the initial and boundary conditions of the equation. The data set can be similar to the KdV One-soliton discussion through the pseudo-spectrum. We found that the neural network with four different combinations of conservation laws predicts soliton interaction and soliton molecule, and the combination of $L3$ has the best prediction result. It can be seen from Table 3 that in the prediction of the interaction of the two soliton, we found that the prediction result of $L3$ are not much different from the classic PINN. For the prediction of soliton molecule, we found that the prediction result of $L3$ is better than the classical result. At the same time, it is found that the loss of $L1$ is smaller than the loss of $L5$, but the prediction result is not as good as that of $L5$. The reason is that there have the energy and momentum constraint equations in the combination of $L1$, but they are missing in the combination of $L5$, that is, the combinations of $L1$ and $L5$ are not the same type of loss.

**Table.3.** Predicted results of two-soliton for KdV equation with different loss combinations

| Soliton type | Different parameters | Loss type | | | | |
|---|---|---|---|---|---|---|
| | | L1 | L2 | L3 | L4 | L5 |
| Soliton interaction | $t|\hat{Q}-Q|_1/Q$ | 0.095279 | 0.316740 | 0.066259 | 0.069018 | 0.014017 |
| | loss | $4.893\times10^{-4}$ | $3.668\times10^{-4}$ | $2.010\times10^{-4}$ | $2.361\times10^{-4}$ | $1.206\times10^{-5}$ |
| Soliton molecule | $t|\hat{Q}-Q|_1/Q$ | 0.026965 | 0.248771 | 0.008644 | 0.009645 | 0.015984 |
| | loss | $8.144\times10^{-6}$ | $7.966\times10^{-6}$ | $2.114\times10^{-6}$ | $2.434\times10^{-6}$ | $1.025\times10^{-5}$ |

Figs. 7(a) and 7(c) show that the complicated nonlinear behaviors of the two-soliton interaction and two-soliton molecule of the KdV equation can still be accurately predicted by the CLCNN method without boundary information. From Figs. 7(b) and Fig. 7(d), with the increase of time, the errors for the loss combinations $L3$ and $L5$ both increase, and this increasing tendency of error for the predicted solution is more obvious for $L5$ used in the traditional PINN method. This further shows that the stability of predicted solutions by the traditional PINN is worse

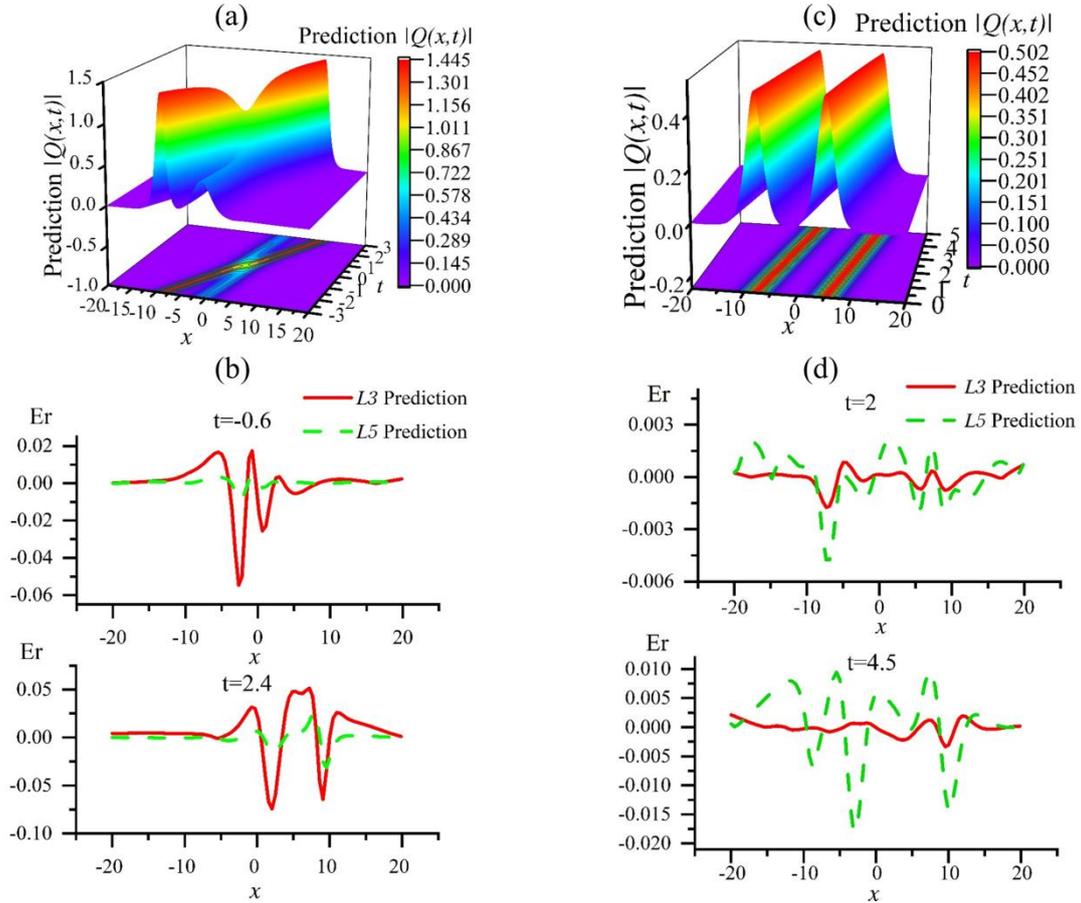

**Fig.7.** Two-soliton solution $Q(x,t)$ of (17). Predicted results of (a) two interacting solitons and (c) two-soliton molecule, comparison between exact and predicted solutions of (b) two interacting solitons and (d) two-soliton molecule at different time. The red line and green dash line denote the predicted solutions with the combinations $L3$ and $L5$ respectively.

## 5. Prediction of soliton solutions of mKdV equation

The mKdV equation has related applications in hydrodynamics and plasma physics, and can be used to describe the kink density wave of traffic jam in hydrodynamics[39]. In plasma physics, it can be used to describe dust acoustic solitary waves in dusty plasma[40]. It can be regarded as a cubic nonlinear KdV equation and an integrable model. Similar to KdV equation, it has a rich family of soliton solutions [41,42]. The specific form of mKdV equation is as follows

$$Q_t + 6Q^2 Q_x + Q_{xxx} = 0, \tag{18}$$

where $Q$ represents the amplitude of the solitary wave, and the subscript $x,t$ denote the normalized distance and time coordinate

Starting from Eq. (18), we construct the corresponding Darboux transformation by using the Lax pair of integrable theory, and we can deduce its energy conservation and momentum conservation laws [35] as follows

$$\begin{aligned} f_{EC} &:= 4Q^3(Q_t + 6Q^2 Q_x + Q_{xxx}) + 2Q_x(-Q_{tx} - 6Q^2 Q_{xx} - 12Q_x^2 Q - Q_{xxxx}), \\ f_{MC} &:= (Q^2)_t + (2QQ_{xx} - Q_x^2 + 3Q^4)_x. \end{aligned} \tag{19}$$

Next, we will consider the case of $L1, L2, L3, L4, L5$ to predict soliton solutions of the mKdV equation, which are the same as that of NLSE in the section 3. Similar to the KdV equation, the result of the loss combination $L3$ is the best in the prediction for solution of mKdV equation, and its loss form is

$$loss = \frac{1}{N_0} \sum_{i=1}^{N_0} |r(x^i,t^i) - r^i|^2 + \frac{1}{N_f} \sum_{j=1}^{N_f} (|f_r(x^j,t^j)|^2 + |f_{MCr}(x^j,t^j)|^2). \tag{20}$$

### 5.1. One-soliton solution

The exact one-soliton solution [38] of Eq. (18) is

$$Q(x,t) = \text{sech}(x-t), x \in [-20, 20], t \in [0,5]. \tag{21}$$

All operation is similar to one-soliton solution of KdV equation in the section 4.1. In the training process, the Latin sampling method is used to only obtain the initial configuration points, and the data of the boundary condition are not used.

Fig. 8 shows the prediction results of one-soliton solution for mKdV equation. Fig. 8 (a) shows the dynamic behavior of one-soliton solution reconstructed by the CLCNN method without boundary information. Fig. 8 (b) displays detailed comparison between the predicted solution and exact solution at $t=1, t=2.5, t=4.5$. Fig. 8 (c) exhibits the gap between the exact solution and the predicted solution. In the whole spatio-temporal region, we find that the gap between the predict value from the CLCNN method and the exact value is very small, and the maximum value is only -0.030. Fig. 8 (d) presents the comparison of neural network prediction for one-soliton solution between four different conservation-law combinations and the loss function combination $L5$ used in the traditional PINN method. Similar to the KdV equation, the prediction of one-soliton solution by the loss function combination $L3$ with the momentum conservation law, the original equation and the initial condition performs best with the relative error as $0.01757$, while the prediction by the loss function combination $L2$ with the momentum and energy conservation laws and the initial condition performs worst with the relative error as $0.06712$.

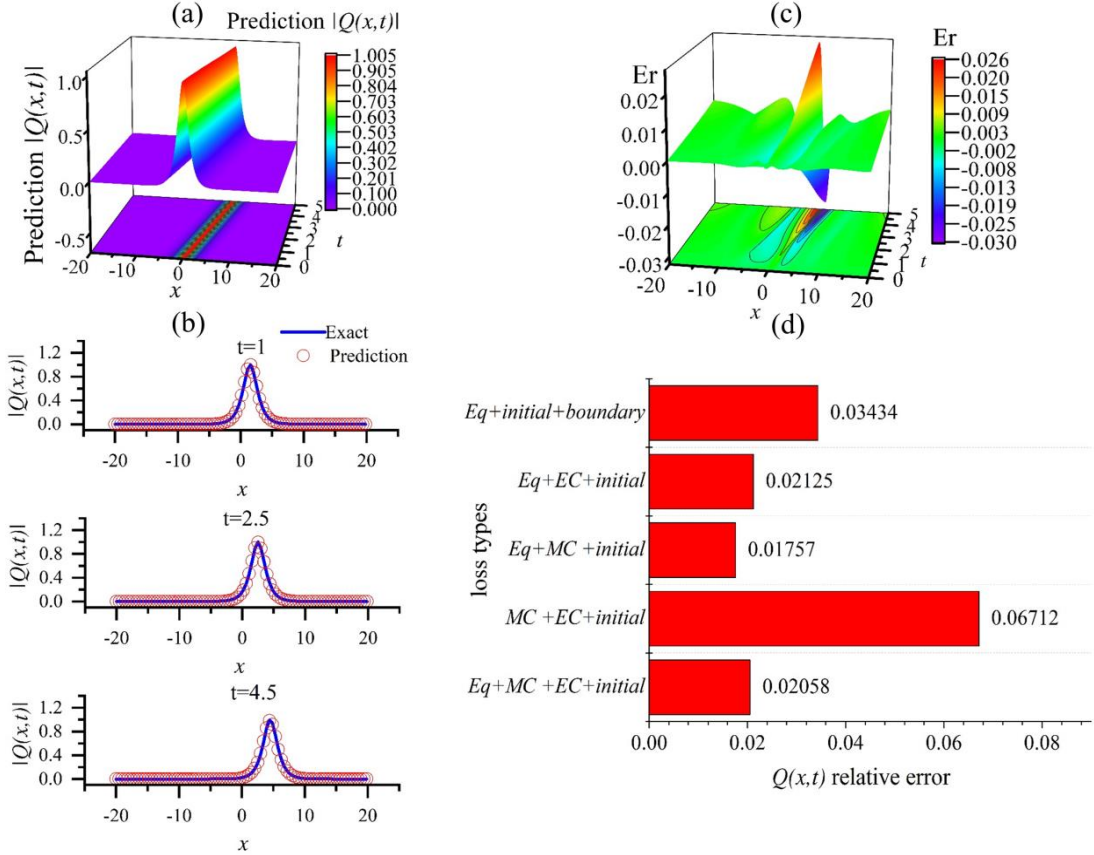

**Fig.8.** One-soliton solution $Q(x,t)$ of (21). (a) Prediction result, (b) comparison between exact and predicted results at different time, (c) diagram of error between exact and predicted solutions, and (d) relative errors of the predicted solution for different loss combinations.

### 5.2. Two-soliton molecule solution

The exact two-soliton molecule solution [43] of Eq. (18) is

$$Q(x,t) = \text{sech}(x-t+4) + \text{sech}(x-t-4), x \in [-15,15], t \in [0,3], \qquad (22)$$

All operation is similar to two-soliton molecule solution of KdV equation in the section 4.2. In the training process, the Latin sampling method is used to only obtain the initial configuration points, and the data of the boundary condition are not used.

Table.4. Relative error of two-soliton molecule with different loss combinations

| Loss combination | L1 | L2 | L3 | L4 | L5 |
|---|---|---|---|---|---|
| Relative error | 0.027180 | 0.066714 | 0.006351 | 0.009333 | 0.008945 |

Table 4 shows the comparison of the relative errors of two-soliton molecule for mKdV equation predicted by different types of loss. From Table 4, we find that the $L3$ prediction with momentum conservation is the best and the relative error of the solution is the smallest, the predicted result for mKdV equation is same as that for KdV equation in Table 3.

Fig. 9(a) displays dynamical behavior of two-soliton molecule reconstructed by the CLCNN method without boundary information. Fig. 9(b) shows the detailed comparison between the predicted solution with loss combinations $L3, L5$ and exact solution at $t=1.2, t=2.7$. From the insets in Fig. 9(b), the prediction by the loss combination $L3$ without boundary information is better than the loss function combination $L5$ used in the traditional PINN method. Fig. 9 (c) displays the gap between the exact solution and the predicted solution. In the whole space-time region, the predicted solution and exact solution can fit well. Fig. 9 (d) shows the loss curve of the neural network with four different loss combinations. We find that the convergence value of the loss curve is the smallest for the loss function combination $L5$ used in the traditional PINN method, but the prediction result is not the best. That's because $L5$ has no conservation law equation constraints, which is different from other types of loss.

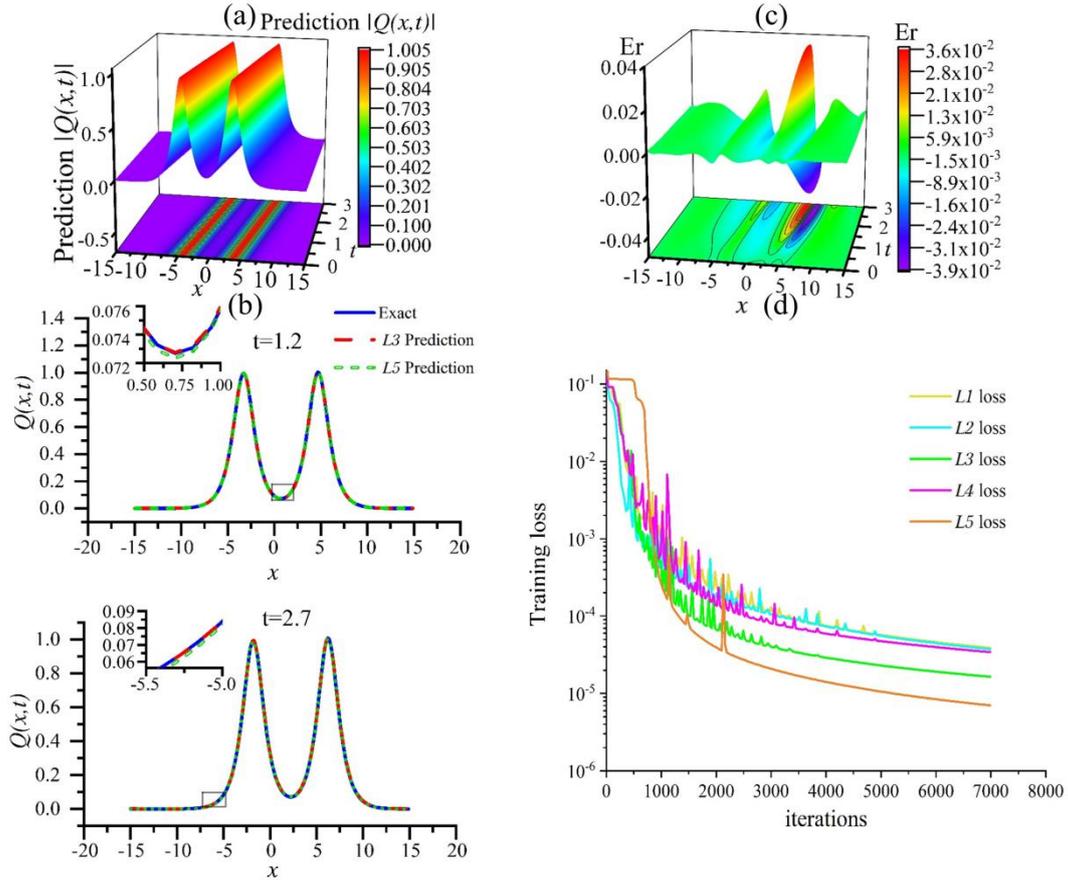

**Fig.9.** Soliton molecule solution $Q(x,t)$ of (22). (a) Prediction result, (b) comparison between exact and predicted results at different time, (c) diagram of error between exact and predicted solutions, and (d) the loss curve of the neural network with four different loss combinations.

## 6. Parameter prediction of physical model

In this section, we will consider the parameter prediction for some NPDEs, such as data-driven NLSE, KdV and mKdV equaitons. The first example, we study a complex NPDE, namely NLSE in the form

$$iQ_t + \lambda_1 Q_{xx} + \lambda_2 |Q|^2 Q = 0, \qquad (23)$$

where the slowly varying envelope $Q$ contains a real part $r$ and an imaginary part $m$, and the coefficients $\lambda_1, \lambda_2$ are the unknown dispersion and nonlinearity coefficients to be trained by the CLCNN method.

Separating real and imaginary parts of Eq.(23) gets

$$\begin{aligned} f_r &:= r_t + \lambda_1 m_{xx} + \lambda_2 (r^2 + m^2) m, \\ f_m &:= m_t - \lambda_1 r_{xx} - \lambda_2 (r^2 + m^2) r. \end{aligned} \qquad (24)$$

In the inverse problem, due to need to predict the coefficients of the NPDE, the loss function combination of conservation-law constraints, must include the original equation. Therefore, we will consider the following loss combinations to train the neural network

$$\begin{aligned} L1 &= Eq + MC + EC, \\ L2 &= Eq + MC, \\ L3 &= Eq + EC, \\ L4 &= Eq, \end{aligned} \qquad (25)$$

where $L4$ is the physical information constraint of the traditional PINN method.

We obtain the unknown parameters of the equation by minimizing the mean square error of the related constraints and the corresponding configuration points. After 10000 times of training, the results show that the performance of the CLCNN method is still very stable, and the relative error of the predicted unknown parameters is smaller than that predicted by the traditional PINN method.

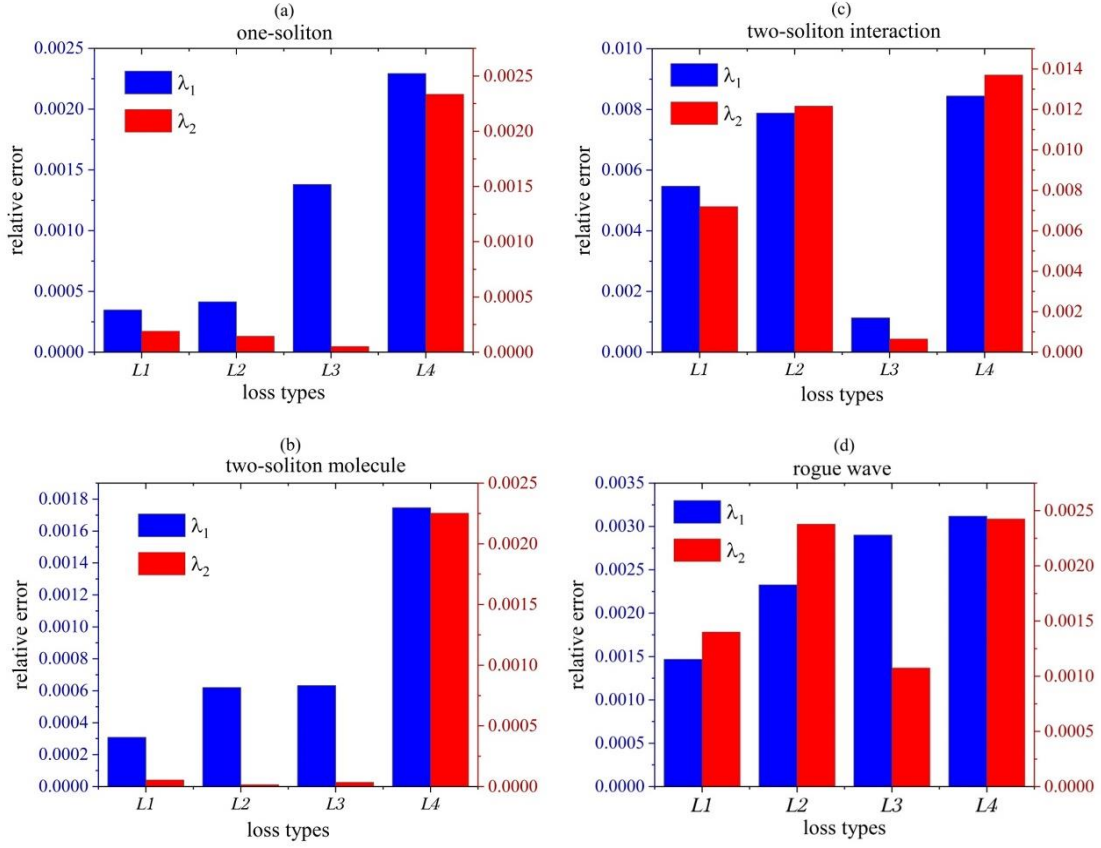

**Fig.10.** The relative errors of parameters $\lambda_1, \lambda_2$ for NLSE predicted by the CLCNN method with different data sets of (a) one-soliton, (b) two-soliton molecule, (c) two-soliton interaction and (d) rogue wave.

From Fig. 10, we find that no matter what kind of data sets including one-soliton in Fig. 10(a), two-soliton molecule in Fig. 10(b), two-soliton interaction in Fig. 10(c) and rogue wave in Fig. 10(d), and no matter what conservation-law constraints as L1, L2 and L3 is added to the neural network, the accuracy of equation parameters predicted by the CLCNN method is better than that of the traditional PINN method. Moreover, for simple one-soliton and two-soliton molecule, we find that the prediction results from the CLCNN method are best, and the accuracy of the parameters from the CLCNN method is much higher than that from the traditional PINN method. For the two-soliton interaction, the prediction results from the CLCNN method with energy conservation law is best, and the accuracy of the parameters from the CLCNN method is much higher than that from the traditional PINN method. For the rogue wave, we find that the prediction result from the CLCNN method with the energy conservation and momentum conservation laws is most accurate. The experiment further shows that the CLCNN method is also effective in the inverse problem to predict the parameters of complex equation.

Next, we study the real NPDEs, namely the KdV and mKdV equations respectively as

$$Q_t + \lambda_1 Q Q_x + \lambda_2 Q_{xxx} = 0,$$
$$Q_t + \lambda_1 Q^2 Q_x + \lambda_2 Q_{xxx} = 0,$$
(26)

where $Q$ represents the solution of the real equation, and the variable $\lambda_1, \lambda_2$ are the unknown nonlinearity and dispersion coefficients to be trained by the CLCNN method.

**Table.5.** Comparison of different identified KdV equations obtained by the CLCNN method with the correct KdV equation

| Item KdV | Dataset type | Loss | Equation | Relative error | |
|---|---|---|---|---|---|
| | | | | $\lambda_1$ | $\lambda_2$ |
| Correct equation | | | $Q_t + 6QQ_x + Q_{xxx} = 0$ | | |
| Identified equations | one-soliton | L1 | $Q_t + 6.00071QQ_x + 1.00021Q_{xxx} = 0$ | 0.01215% | 0.01155% |
| | | L2 | $Q_t + 6.00130QQ_x + 1.00040Q_{xxx} = 0$ | 0.01939% | 0.05772% |
| | | L3 | $Q_t + 5.99850QQ_x + 0.99926Q_{xxx} = 0$ | 0.02993% | 0.07513% |
| | | L4 | $Q_t + 5.96147QQ_x + 0.98569Q_{xxx} = 0$ | 0.62601% | 1.41114% |
| | two-soliton molecule | L1 | $Q_t + 5.99907QQ_x + 0.99980Q_{xxx} = 0$ | 0.01628% | 0.01928% |
| | | L2 | $Q_t + 5.99812QQ_x + 0.99968Q_{xxx} = 0$ | 0.03124% | 0.03355% |
| | | L3 | $Q_t + 5.99792QQ_x + 0.99914Q_{xxx} = 0$ | 0.03533% | 0.07936% |
| | | L4 | $Q_t + 5.97886QQ_x + 0.99219Q_{xxx} = 0$ | 0.34479% | 0.75708% |
| | two-soliton interaction | L1 | $Q_t + 6.00465QQ_x + 1.00107Q_{xxx} = 0$ | 0.07746% | 0.10691% |
| | | L2 | $Q_t + 6.01980QQ_x + 1.00470Q_{xxx} = 0$ | 0.32922% | 0.46905% |
| | | L3 | $Q_t + 6.01149QQ_x + 1.00329Q_{xxx} = 0$ | 0.19089% | 0.32829% |
| | | L4 | $Q_t + 5.96446QQ_x + 0.98655Q_{xxx} = 0$ | 0.58544% | 1.33091% |

**Table.6.** Comparison of different identified mKdV equations obtained by the CLCNN method with the correct mKdV equation

| Item mKdV | Dataset type | Loss | Equation | Relative error $\lambda_1$ | Relative error $\lambda_2$ |
|---|---|---|---|---|---|
| Correct equation | | | $Q_t + 6Q^2 Q_x + Q_{xxx} = 0$ | | |
| Identified equations | one-soliton | L1 | $Q_t + 5.99629 Q^2 Q_x + 0.99916 Q_{xxx} = 0$ | 0.08376% | 0.06188% |
| | | L2 | $Q_t + 5.99790 Q^2 Q_x + 0.99952 Q_{xxx} = 0$ | 0.03490% | 0.04784% |
| | | L3 | $Q_t + 5.99450 Q^2 Q_x + 0.99882 Q_{xxx} = 0$ | 0.09154% | 0.11744% |
| | | L4 | $Q_t + 5.63551 Q^2 Q_x + 0.90756 Q_{xxx} = 0$ | 6.02932% | 9.17620% |
| | two-soliton molecule | L1 | $Q_t + 6.02260 Q^2 Q_x + 1.00510 Q_{xxx} = 0$ | 0.37770% | 0.47556% |
| | | L2 | $Q_t + 6.00030 Q^2 Q_x + 1.00006 Q_{xxx} = 0$ | 0.00449% | 0.00634% |
| | | L3 | $Q_t + 5.98276 Q^2 Q_x + 0.99606 Q_{xxx} = 0$ | 0.28436% | 0.39175% |
| | | L4 | $Q_t + 5.90562 Q^2 Q_x + 0.97281 Q_{xxx} = 0$ | 1.57563% | 2.68242% |

In Tables 5 and 6, we take different types of soliton solutions as data sets, and observe the training results and relative errors of unknown coefficients in the case of $L1, L2, L3, L4$. No matter what conservation-law constraints as $L1, L2$, and $L3$ is added to the neural network, the accuracy of equation parameters predicted by the CLCNN method is better than that of the traditional PINN method, which is consistent with the result in NLSE. For different data sets, in the KdV equation, the relative error of parameters predicted by CLCNN with energy and momentum conservation laws is the smallest. In the mKdV equation, CLCNN with momentum conservation law has the highest accuracy in predicting the equation parameters. It further shows that the CLCNN method is still effective in the inverse problem to predict the parameters of real NPDEs.

7. **Conclusion**

In this paper, we first propose the CLCNN method to predict soliton solutions of NPDEs without the boundary information. This is achieved by deriving integrable information such as conservation laws from the equation itself, integrating more integrable information into the neural network, and forming more severe constraints on the network output. Similar to the PINN method, our proposed method can be used to solve any integrable NPDEs. For KdV and mKdV equations, the best loss combination to predict all soliton solutions is the form $L_3$, namely, the loss combination of the original equation and the momentum conservation law. For the NLSE, we find that the best loss combination to predict all soliton solutions is the form $L_4$, that is, the loss combination of the original equation and the energy conservation law. These predictions to various soliton solutions of the NLSE, KdV and mKdV equations are all better than results from the traditional PINN. The above examples also prove that incorporating more conservation-law constraint information can well replace the role of boundary condition and realize the prediction of solutions for NPDEs without the boundary information.

In the inverse problem, we find that the parameter accuracy of the CLCNN method is much higher than that of the traditional PINN method. Results from the forward and inverse problems of the physical models show that the CLCNN method has the advantages of high accuracy and wide applicability. Moreover, the variable learning rate introduced in this paper can improve the convergence speed and is not easy to fall into local minimum.

Although we have obtained excellent results, we still need to continue to study some problems, including the optimization of the neural network and combination of these results in this paper with experiments.

## Acknowledgements

This work is supported by the Zhejiang Provincial Natural Science Foundation of China (Grant No. LR20A050001) and the National Natural Science Foundation of China (Grant Nos. 12075210 and 11874324).

## Conflict of interest

The authors have declared that no conflict of interest exists.

## Ethical Standards

This Research does not involve Human Participants and/or Animals.

## References
[1] Draper L. 'Freak' ocean, Mar. Obs. 1965; 35:193-195.
[2] Mandal D, Sharma D. Nonlinearly interacting trapped particle solitons in collisionless plasmas. Phys. Plasmas 2016;23(2):022108.
[3] Tabbert F, Gurevich SV, Panajotov K, Tlidi M. Oscillatory motion of dissipative solitons induced by delay-feedback in inhomogeneous Kerr resonators. Chaos Soliton. Fract. 2021;152:111317.
[4] Nandy S, Barthakur A. Dark-bright soliton interactions in coupled nonautonomous nonlinear Schrodinger equation with complex potentials. Chaos Soliton. Fract. 2021;143:110560.
[5] Bansal A, Kara AH, Biswas A, Moshokoa SR, Belic M. Optical soliton perturbation, group invariants and conservation laws of perturbed Fokas-Lenells equation. Chaos Soliton. Fract. 2018;114:275-280.

[6] Noether E. Invariant Variational Problems. In: The Noether Theorems. Sources and Studies in the History of Mathematics and Physical Sciences, New York, NY, 2011.

[7] Frasca-Caccia G, Hydon PE. Numerical preservation of multiple local conservation laws. Appl. Math. Comput. 2021;403:126203.

[8] Matsukidaira J, Sasuma J, Strampp W. Conserved quantities and symmetries of KP hierarchy. J. Math. Phys. 1990;31:1426.

[9] Wu S, Roberts K, Datta S, Du JC, et al.. Deep learning in clinical natural language processing: a methodical review. J. Am. Med. Inform. Assn. 2020;27:457-470.

[10] Wu GZ, Fang Y, Wang YY, Wu GC, Dai CQ. Predicting the dynamic process and model parameters of the vector optical solitons in birefringent fibers via the modified PINN. Chaos Soliton. Fract. 2021;152: 111393.

[11] Surazhevsky IA, Demin VA, Ilyasov AI, et al.. Noise-assisted persistence and recovery of memory state in a memristive spiking neuromorphic network. Chaos Soliton. Fract. 2021;146:14.

[12] Zhu YH, Zabaras N, Bayesian deep convolutional encoder–decoder networks for surrogate modeling and uncertainty quantification. J.Comput. Phys. 2018;366:415–447.

[13] Scarselli F, Tsoi AC. Universal approximation using feedforward neural networks: A survey of some existing methods, and some new results. Neural Netw.1998;11(1):15–37.

[14] Hutzenthaler M, Jentzen A, Kruse T, et al.. Overcoming the curse of dimensionality in the numerical approximation of semilinear parabolic nonlinear partial differential equations. P. Roy. Soc. A-Math. Phy. 2020;476 (2244):20190630.

[15] Grohs P, Hornung F, Jentzen A, Wurstemberger PV. A proof that artificial neural networks overcome the curse of dimensionality in the numerical approximation of black-scholes nonlinear partial differential equations. arXiv preprint arXiv:1809.02362.

[16] Hutzenthaler M, Jentzen A, Kruse T, Nguyen TA. A proof that rectified deep neural networks overcome the curse of dimensionality in the numerical approximation of semilinear heat equations. arXiv preprint arXiv:1901.10854.

[17] Lagaris IE, Likas A, Fotiadis DI. Artificial neural networks for solving ordinary and nonlinear partial differential equations, IEEE Trans. Neural Netw. 1998;9(5):987–1000.

[18] Lagaris IE, Likas AC, Papageorgiou DG. Neural-network methods for boundary value problems with irregular boundaries, IEEE Trans. Neural Netw.2000;11(5):1041–1049.

[19] Raissi M, Perdikaris P, Karniadakis GE. Physics-informed neural networks: A deep learning framework for solving forward and inverse problems involving nonlinear partial differential equations, J. Comput. Phys. 2019;378:686-707.

[20] Zhou ZJ, Yan ZY. Solving forward and inverse problems of the logarithmic nonlinear Schrödinger equation with PT-symmetric harmonic potential via deep learning. Phys. Lett. A 2021; 387:127010.

[21] Wang L, Yan ZY. Data-driven rogue waves and parameter discovery in the defocusing nonlinear Schrödinger equation with a potential using the PINN deep learning. Phys. Lett. A 2021;404:127408.

[22] Li J, Chen Y. A deep learning method for solving third-order nonlinear evolution equations. Commun. Theor. Phys. 2020 ;72(11):115003.

[23] Jagtap AD, Kharazmi E, Karniadakis GE. Conservative physics-informed neural networks on discrete domains for conservation laws: Applications to forward and inverse problems. Comput. Method. Appl. M. 2020;365(15):113028.

[24] Raissi M, Yazdani A, Karniadakis GE. Hidden fluid mechanics: A navier-stokes informed deep learning framework for assimilating flow visualization. Science 2020;367(6481):1026-1030.

[25] Sun LN, Gao H, Pan SW, et al.. Surrogate modeling for fluid flows based on physics-constrained deep learning without simulation data. Comput. Method. Appl. M. 2019;361(1):112732.

[26] Lin S, Chen Y. A two-stage physics-informed neural network method based on conserved quantities and applications in localized wave solutions. arXiv preprint arXiv: 2107.01009.

[27] Cousins W, Sapsis TP. Unsteady evolution of localized unidirectional deep-water wave groups, Phys. Rev. E 2015;91(6):063204

[28] Agrawal GP. Nonlinear fiber optics, in: Nonlinear Science at the Dawn of the 21st Century 2000; pp. 18-36.

[29] Parkins A, Walls D. The physics of trapped dilute-gas Bose-Einstein condensates. Phys. Rep. 1998;303(1):1-80.

[30] Li Y. Soliton and Integrable System, in: Shanghai Scientific and Technological Education Publishing House 1999; pp.140-142.

[31] Pu JC, Li J, Chen Y. Soliton, breather, and rogue wave solutions for solving the nonlinear Schrödinger equation using a deep learning method with physical constraints. Chinese Phys. B 2021;30(6):060202.

[32] McKay MD, Beckman RJ, Conover WJ. A Comparison of Three Methods for Selecting Values of Input Variables in the Analysis of Output from a Computer Code. Technometrics. 1979; 21(2):239–245.

[33] Khalique CM. Closed-form solutions and conservation laws of a generalized Hirota-Satsuma coupled KdV system of fluid mechanics. Open Phys. 2021;19(1):18-25.

[34] Mottaghizadeh M, Eslami P. Cylindrical and spherical ion-acoustic solitons in electron-positive ion-negative ion plasmas. Indian. J. Phys. 2012;86(1):71-75.

[35] Miki W, Heiji S, Kimiaki K. Relationships among Inverse Method, Backlund, Transformation and an Infinite Number of Conservation Laws. Prog. Theor. Phys. 1975;53(2):419.

[36] Korteweg DI, Vries GD. On the change of form of long waves advancing in a rectangular canal and on a new type of long stationary waves. Phil. Mag. 1895;39(240):422-443.

[37] Miura RM, Gardner CS, Kruskal MD. Korteweg-de Vries Equations and Generalizations. II. Existence of Conservation Laws and Constants of Motion. J. Math. Phys. 1968;9(8):1204-1209.

[38] Li J, Chen Y. A deep learning method for solving third-order nonlinear evolution equations Learning, Commun. Theor. Phys. 2020;72(10):115003.

[39] Alejo MA, Muoz C. Nonlinear Stability of MKdV Breathers. Commun. Math. Phys. 2013;324(1):233–262.

[40] Lei Y, Shi Z, Zhou B. Kink–antikink density wave of an extended car-following model in a cooperative driving system. Commun. Nonlinear Sci. 2018;13(10):2167-2176.

[41] Fu Z, Liu S, Liu S. New solutions to mKdV equation. Phys. Lett. A 2004;326:364–374.

[42] Seadawy AR, Iqbal M, Lu DC. Propagation of kink and anti-kink wave solitons for the nonlinear damped modified Korteweg-de Vries equation arising in ion-acoustic wave in an unmagnetized collisional dusty plasma. J. Am. Med. Inform. Assn. 2020;544:123560.

[43] Chai YZ, Jia TT, Hao HQ, Zhang JW. Exp-Function Method for a Generalized MKdV Equation. Discrete Dyn. Nat. Soc. 2014;2014:153974.